\documentclass[aps,pra,twocolumn,showpacs,preprintnumbers,amsmath,amssymb,
floatfix,superscriptaddress]{revtex4-1}

\usepackage{graphics}
\usepackage{graphicx}
\usepackage{amsmath}
\usepackage{setspace}
\usepackage{braket}
\usepackage{epstopdf}
\usepackage{comment}
\usepackage{hyperref}
\usepackage{bm}
\usepackage{xfrac}

\hypersetup{%
   pdfpagemode=None, 
   pdfstartpage=1,
   pdfmenubar=true,
   pdftoolbar=true,
   colorlinks = true,
   linkcolor=blue,
   citecolor=blue,
   urlcolor=blue,
   bookmarksopen=false
 }

\newcommand{\affA}{Van der Waals-Zeeman Institute, Institute of Physics,
University of Amsterdam, 1098 XH Amsterdam, Netherlands}
\newcommand{\affB}{Institute for Experimental Physics, University of Innsbruck, 6020, Innsbruck, Austria}
\newcommand{\affC}{Physikalisch-Technische Bundesanstalt, Bundesallee 100, 38116 Braunschweig, Germany}

\begin{document}

\title{Experimental setup for studying an ultracold mixture of trapped Yb$^+$-$^6$Li}

\author{H.~Hirzler}\affiliation{\affA}
\author{T.~Feldker}\affiliation{\affB}
\author{H.~F\"urst}\affiliation{\affC}
\author{N.\,V.~Ewald}\affiliation{\affA}
\author{E.~Trimby}\affiliation{\affA}
\author{R.\,S.~Lous}\affiliation{\affA}
\author{J. Arias Espinoza}\affiliation{\affA}
\author{M.~Mazzanti}\affiliation{\affA}
\author{J.~Joger}\affiliation{\affA}

\author{R.~Gerritsma}\affiliation{\affA}

\date{\today}

\begin{abstract}
We describe and characterize an experimental apparatus that has been used to study interactions between ultracold lithium atoms and ytterbium ions. The preparation of ultracold clouds of Li atoms is described as well as their subsequent transport and overlap with Yb$^+$ ions trapped in a Paul trap. We show how the kinetic energy of the ion after interacting with the atoms can be obtained by laser spectroscopy. From analyzing the dynamics of the ion in the absence of atoms, we conclude that background heating, due to electric field noise, limits attainable buffer gas cooling temperatures. We suspect that this effect can be mitigated by noise reduction and by increasing the density of the Li gas, in order to improve its cooling power. Imperfections in the Paul trap lead to so-called excess micromotion, which poses another limitation to the buffer gas cooling. We describe in detail how we measure and subsequently minimize excess micromotion in our setup. We measure the effect of excess micromotion on attainable ion temperatures after buffer gas cooling and compare this to molecular dynamics simulations which describe the observed data very well. 
\end{abstract}

\maketitle
\section{Introduction}

In recent years, a new field in cold atomic physics has developed in which laser-cooled trapped ions are merged with ultracold atomic gases~\cite{Harter:2013,Cote:2016,Tomza:2017cold}. These hybrid systems have been used for studying cold chemistry and collisions between ions and atoms~\cite{Zipkes:2010,Schmid:2010,Ratschbacher:2012,Ratschbacher:2013,Haze:2015,Meir:2016,Fuerst:2018:spin,Ewald:2019}. Furthermore, a range of applications in the quantum regime have been suggested. These include the possibility to use degenerate clouds of atoms as a coolant for trapped ions, studying the quantum many-body physics of interacting clouds of atoms and ion crystals~\cite{Bissbort:2013,Negretti:2014}, using trapped ions as local field probes in atomic quantum systems~\cite{Kollath:2007} as well as applications in quantum information processing~\cite{Doerk:2010,Secker:2016}. However, it was realized that the dynamic electric fields of a Paul trap used for trapping the ions cause significant problems when combining them with cold atoms. In particular, it was found that energy can be extracted from the trap in an atom-ion collision, which causes the ion to acquire significantly larger energies than that of the atomic cloud~\cite{Cetina:2012,Chen:2014,Rouse:2017,Meir:2016}. It was pointed out, that reaching a regime in which the collisional angular momentum is quantized (the so called $s$-wave or quantum regime) would only be possible for the largest available mass ratios between the ion and atom species~\cite{Cetina:2012,Fuerst:2018,Haze:2018,Pinkas:2020}. Combined with the benefit of straightforward laser (pre-)cooling and manipulation, $^6$Li/Yb$^+$ stands out as a particularly promising species combination. Very recently, we have observed collision energies at around the $s$-wave limit for the first time in this mixture~\cite{Feldker:2020}. We also note that a promising alternative solution is to use optical potentials for the trapped ion~\cite{Schmidt:2020,Perego:2020}.

Although both, $^6$Li and Yb$^+$ are routinely trapped and manipulated in atomic physics laboratories, combining the two is not straightforward technically as well as conceptually. {In particular, when $^6$Li is loaded from an atomic beam into a magneto-optical trap~\cite{Hulet:2020, Murthy:2019}, care needs to be taken that the atoms do not contaminate the electrodes used for trapping the ions. }
Oxidation of deposited Li over time can cause patch potentials that deteriorate the trapping potential of the ions~\cite{Harlander:2010}. Furthermore, preparing ultracold $^6$Li requires high-power lasers for optical trapping and magnetic fields of $\sim$~80~mT to reach the Feshbach resonances needed for efficient evaporative cooling, demands that are not straightforwardly combined with ion trapping. Finally, neither ions nor atoms can be laser-cooled while interacting as collisions between atoms and ions in electronically excited states lead to losses~\cite{Joger:2017}. 

In this work, we describe in detail the experimental apparatus that we have built to study interactions between Yb$^+$ and $^6$Li that addresses all the issues mentioned above. The paper is organized as follows: In section~\ref{Sec_Setup} we describe the experimental setup, including our ion trap~\ref{Subsec_ions}, atomic traps~\ref{Subsec_atoms}, and our techniques to overlap them~\ref{Subsec_overlap}. In section~\ref{Sec_spectro} we show how we can measure the average kinetic energy of a trapped ion. We also use these techniques to measure and compensate excess micromotion of our trapped ion, as described in subsection~\ref{Subsec_MMcomp}. In section~\ref{Sec_atomsandions}, we describe experiments that we performed with interacting atoms and ions. By observing the ion dynamics after buffer gas cooling to a temperature of 90(35)~$\mu$K, we extract a heating rate of 85(50)~$\mu$K/s in the radial direction of motion as described in subsection~\ref{Subsec_heatingrates}. In subsection~\ref{Subsec_EMMbuffergascool}, we study the effect of imperfect micromotion compensation on the attainable ion temperatures after buffer gas cooling. Finally, in section~\ref{Sec_conclusions}, we draw conclusions and give an outlook towards future experiments.

\section{Experimental setup}
\label{Sec_Setup}

\begin{figure*}
	\includegraphics[width=0.65\linewidth]{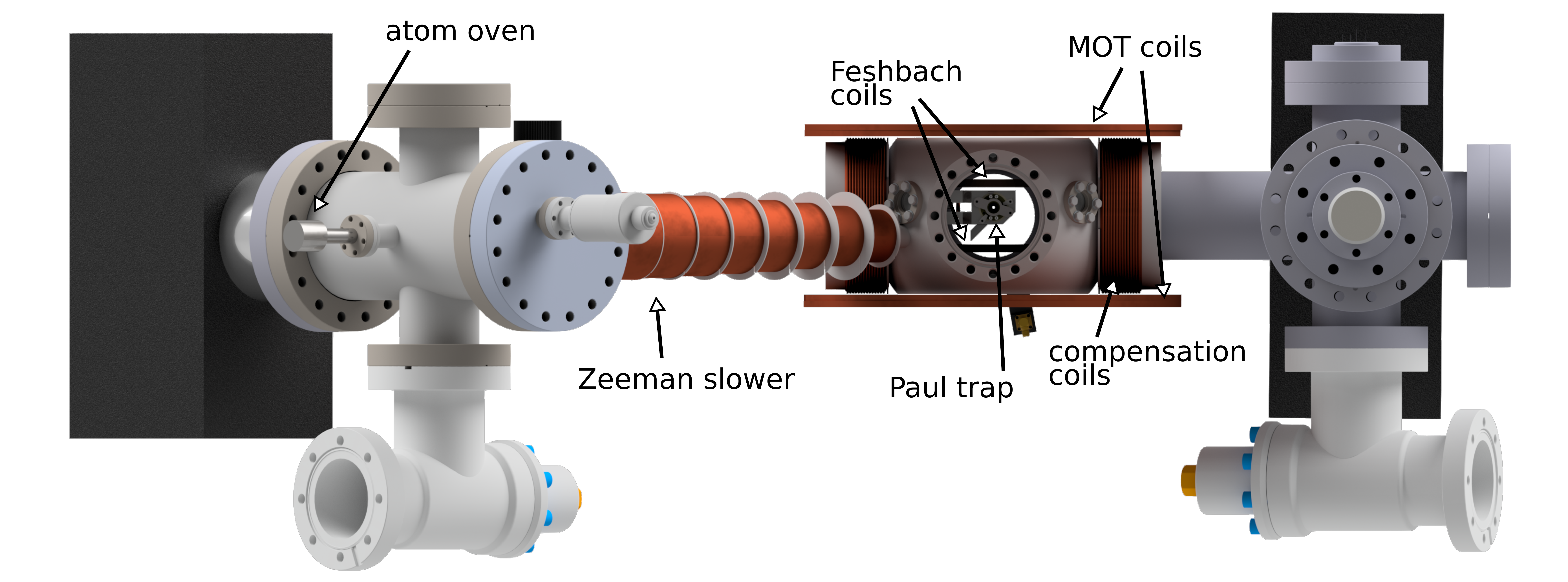}
	\caption{Overview of our experimental apparatus for combining ultracold atoms with trapped ions. Li atoms are heated from an oven and subsequently slowed down in a Zeeman slower towards the main chamber. A 45$^\circ$ mirror is used to create a mMOT about $20\,$mm underneath the Paul trap. The field gradient is provided by a set of MOT coils and a homogeneous offset field is produced with the Feshbach coils and four horizontal compensation coils. A strong magnetic quadrupole field is created with the Feshbach coils to trap and to transport the atoms upwards and into the Paul trap.}
	\label{fig:CAD}
\end{figure*}

In our setup we combine a linear Paul trap, where ytterbium ions are trapped with the help of static and dynamic electric fields, with a cloud of cold lithium atoms trapped in a magnetic quadrupole trap or an optical dipole trap. {In Fig. \ref{fig:CAD} a drawing of our  vacuum system is shown. It consists of a single main chamber including the Paul trap and it is surrounded by magnetic field coils that are necessary for manipulating both atoms and ions.}

\subsection{Trapped ions}
\label{Subsec_ions}

We trap Yb$^+$ ions in a linear Paul trap as indicated in Fig.\,\ref{fig:CAD}. We operate the trap at a drive frequency of about $\Omega_{\text{rf}}=2\pi\times 2$\,MHz with an amplitude of up to $U_{\text{rf}} = 200$\,V. This results in a typical radial potential with trap frequencies of $\omega_{\text{rad}} = 2\pi\times 100-350\,$kHz. A small symmetric offset voltage on two electrodes  lifts the degeneracy of the two radial modes of motion. An axial potential with trap frequencies up to $\omega_\text{ax} =2\pi\times 120$\,kHz is generated by applying voltages of up to $V_{\text{ax}} = 150$\,V to the endcap electrodes. The ions are trapped around 1.5~mm from the radial electrodes and 5~mm from the endcap electrodes.

\begin{figure}
	\includegraphics[width=0.95\columnwidth]{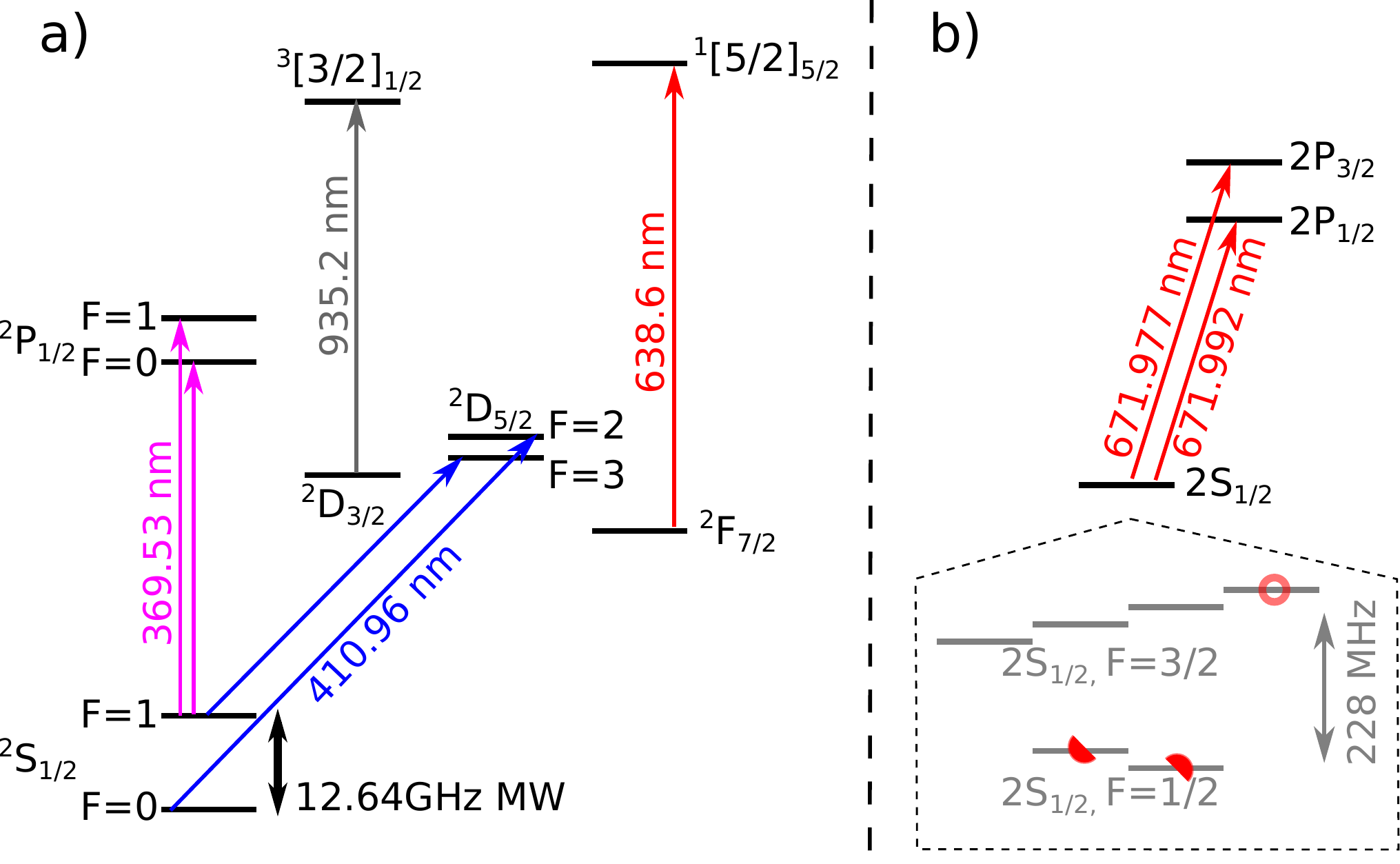}
	\caption{Reduced energy levels for the hybrid atom-ion system. a) Energy levels and transitions of $^{171}$Yb$^+$ used in the experiment. b) Energy levels and transitions of $^6$Li. The inset shows the ground state hyperfine-structure and Zeeman-levels. Atoms are trapped either in the $|F=3/2, m_F=3/2\rangle$ state in the magnetic trap or in a spin mixture of the $|F=1/2, m_F=\pm1/2\rangle$ states in the optical dipole trap. } \label{fig:YbScheme}
\end{figure}

\begin{figure}
	\includegraphics[width=0.95\columnwidth]{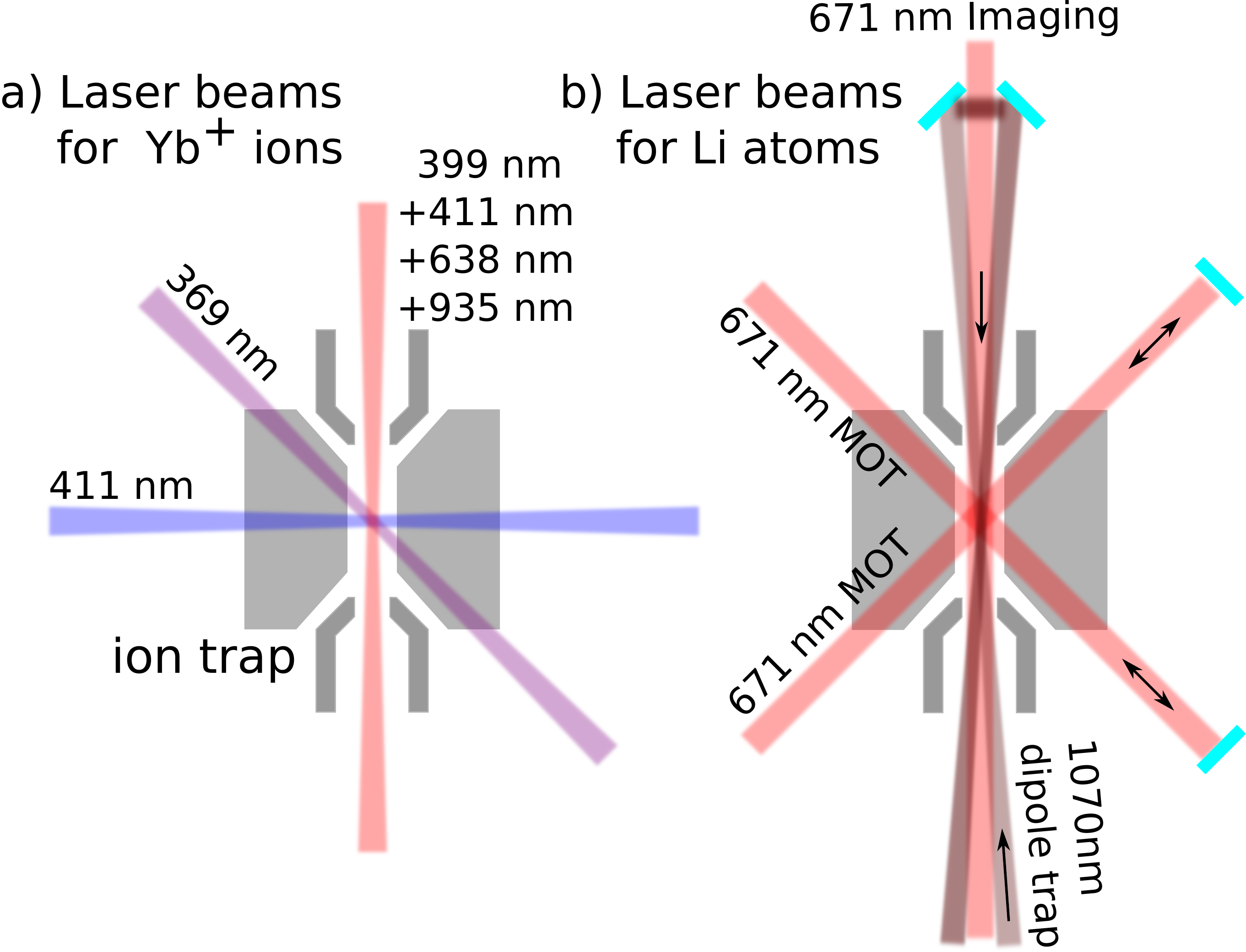}
	\caption{Sketch of the laser beams for the ions and atoms in horizontal cross section centered through the ion trap. a) Laser beams used for trapping, cooling, spectroscopy and state preparation of the ions. The laser beam at 369\,nm wavelength for Doppler cooling and imaging enters at an angle of 45$^\circ$ so that it has a projection onto all motional modes. The repumper beams at wavelengths of 935\,nm and 638\,nm are aligned along the trap axis, so that even for long ion crystals we have a homogeneous illumination of the ions. The photo-ionization beam at 399\,nm wavelength is aligned along the trap axis as well. We align the spectroscopy beam at 411\,nm either parallel to the trap axis or perpendicular to the trap axis with overlap with both radial modes of motion. b) Laser beams used for atom trapping and cooling. Three retro-reflected beams (the vertical beam is omitted in the figure) with circular polarization for a MOT at the position of the dipole trap center. The beams for the optical dipole trap enter through holes in the endcaps at an angle of 5$^\circ$ with respect to the Paul trap axis leading to a cigar-shaped potential. We use a resonant laser beam aligned parallel to the trap axis to perform absorption imaging of the atomic cloud. }
	\label{fig:LaserBeams}
\end{figure}

{Fig.~\ref{fig:YbScheme} a) shows the relevant energy levels and transitions of $^{171}$Yb$^+$ and Fig.~\ref{fig:LaserBeams} a) shows the laser beam orientations with respect to the Paul trap.} We ionize atoms from a thermal beam of neutral ytterbium inside the trap. For isotope selectivity, we use a two-step ionization scheme with wavelengths of 399\,nm and 369\,nm. This allows us to trap any stable Yb$^+$ isotope except $^{173}$Yb$^+$. The ions are Doppler-cooled using the S$_{1/2} \rightarrow \text{P}_{1/2}$ transition at 369\,nm. In the case of $^{171}$Yb$^+$ with hyperfine structure we use the approximately closed $\ket{S_{1/2},F=1}\rightarrow \ket{P_{1/2},F=0}$ transition, as shown in Fig.~\ref{fig:YbScheme} a). We prevent population trapping in the metastable $D_{3/2}$ state by applying light at 935\,nm wavelength to excite the $D_{3/2} \rightarrow$ $[3/2]_{1/2}$ transition. Excitation of the $S_{1/2}\rightarrow D_{5/2}$ transition near $411\,$nm leads to  a subsequent population of the metastable $F_{7/2}$ state (radiative lifetime $t_F \approx 10$ yrs) with 83\% probability~\cite{Taylor:1997}. 
We employ an additional light field of a wavelength near 638\,nm to pump the population from the $F_{7/2}$ state back to the ground state. For detection, we image the fluorescence light during Doppler cooling with a sCMOS camera as well as with a photomultiplier tube (PMT) for fast imaging. For motional spectroscopy, we excite the  $S_{1/2} \rightarrow D_{5/2}$ transition at 411\,nm. The natural linewidth of this quadrupole transition of {$\Gamma = 22\,$Hz}  allows for resolving individual motional sidebands \cite{Roberts:1999}. 
We use an ECDL laser to obtain the light near 411\,nm, which is stabilized to a commercial high Finesse cavity ($\mathcal{F}= 3\times 10^4$) with a high bandwidth PDH loop. The FWHM linewidth of the stabilized laser is about 2\,kHz, deduced from spectroscopy on a magnetic field-insensitive transition.  Two beams of 411~nm light enter the vacuum system. One is aligned along the Paul trap axis and the other vertically. In this way we can either couple to the axial modes of motion of the ion or to the combined radial modes.

We have the option to modulate sidebands at $2\pi\times 2.1$\,GHz on the cooling beam with an electro-optical modulator (EOM). With this additional frequency we can drive the $\ket{S_{1/2},F=1}\rightarrow \ket{P_{1/2},F=1}$ transition in order to optically pump the ion into the $\ket{S_{1/2},F=0}$ state~\cite{Olmschenk:2007}. 

We employ a microwave field to mix the hyperfine ground states of $^{171}$Yb$^+$ when required. We generate this field by mixing the 12.6\,GHz output of a frequency generator with a variable frequency from a versatile frequency generator and amplify the resulting frequency to 10\,W. The microwave field is coupled to the ion via a microwave horn antenna placed outside the vacuum chamber. 
 
\begin{figure}
	\includegraphics[width=0.7\columnwidth]{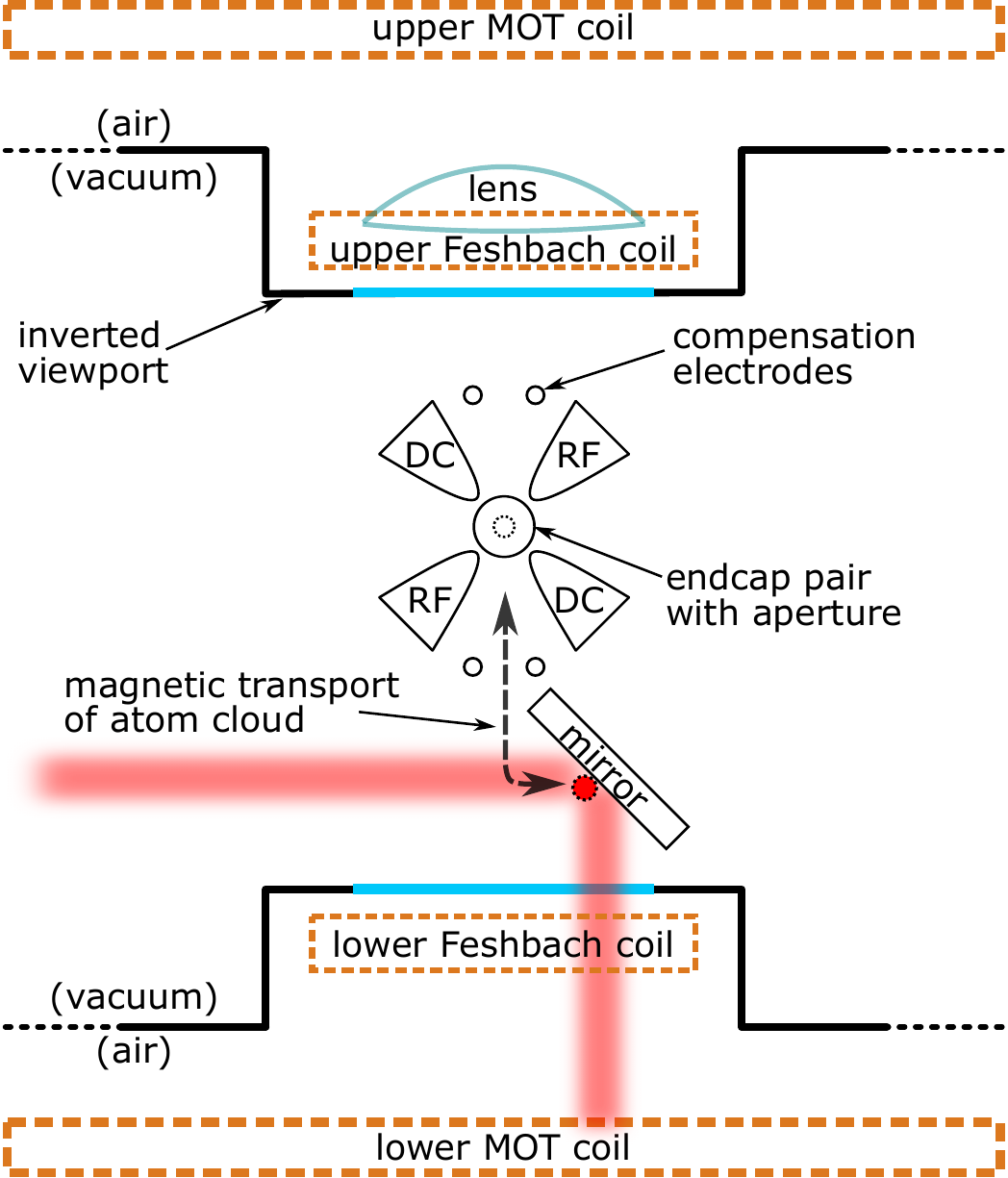}
	\caption{Sketch of the vertical cross section through the experiment. The center of the Paul trap is situated about 20\,mm above the mirror for the mMOT. The atoms are initially trapped in front of the 45$^\circ$ mirror and subsequently magnetically transported up into the ion trap. Two sets of coils labeled MOT coil and Feshbach coil provide magnetic fields {or field gradients, depending on the polarity.}  The lens for ion imaging has a numerical aperture of NA$\lesssim 0.6$ and is placed a few\,cm above the Paul trap center in a re-entrance viewport.}
	\label{fig:SetupVert}
\end{figure}

\subsection{Ultracold atoms}
\label{Subsec_atoms}

The atomic beam comes from the lithium oven and is slowed down using a Zeeman slower, see  Fig. \ref{fig:CAD}. After entering the main chamber, the atoms are trapped in a mirror magneto-optical trap (mMOT) about 20\,mm below the center of the Paul trap as shown in Fig.~\,\ref{fig:SetupVert}. 

Fig. \ref{fig:YbScheme}~b) shows the relevant transitions of $^6$Li. We use red-detuned light  from the D2-line at 670.977\,nm and magnetic field gradients of {$g_z=0.44$\,T/m}. In addition, light shifted by +228\,MHz acts as a repumper and prevents population trapping in the lower hyperfine state $F=1/2$. The mMOT consists of two retro-reflected beams with a beam waist of {$\omega_0$=25}\,mm, a power of 75\,mW, and a detuning of {-34\,MHz} from resonance. 

To generate strong homogeneous B-fields or B-field gradients we have two sets of coils in the vertical direction as depicted in Fig.~\ref{fig:CAD} and Fig.~\ref{fig:SetupVert}. The MOT coils are vertically
aligned 62.5\,mm away from the ion trap center. These coils can be {switched such that the result is either a homogeneous field  (up to $B_0=40$\,mT) or a field gradient (up to $g_z=0.5\,$T/m).} Additionally we have a set of smaller coils (Feshbach coils) closer to the ion trap, positioned inside re-entrance viewports. These coils can be switched on shorter time scales of a few\,ms and provide a field of up to $B_0=80$\,mT or a gradient of up to $g_z=2.8\,$T/m, depending on the polarity. Additionally, compensation coils in the two horizontal directions allow us to apply small offset fields.

We load atoms into the mMOT for 3--10\,s before we compress the mMOT by simultaneously ramping down the laser intensity to 0.5\,mW  and the detuning to -10\,MHz in 3\,ms. Subsequently, we apply a bias field of 0.6\,mT and partially pump the $^6$Li atoms to the magnetically trappable $F=3/2$, $m_F=3/2$ state with a 100$\,\mu$s pulse of circularly polarized light resonant to the D1-line at 670.992\,nm wavelength. The quadrupole field remains on during optical pumping so that the polarized atoms are immediately trapped. Even if we loose a fraction of the atoms due to the non-homogeneous magnetic field, we found this to be the most efficient way of loading the magnetic trap in our setup. In particular, it proved impossible to switch magnetic fields fast enough to perform optical pumping in a homogeneous field. We attribute this effect to the bulky stainless steel ion trap, which prevents fast field switching in its vicinity because of the induced eddy currents.  We end up with up to 10$^8$ magnetically trapped atoms in the state $F=3/2$, $m_F=3/2$ at a temperature of about $T_\text{atom}=300\,\mu$K.

\begin{figure}
	\includegraphics[width=\columnwidth]{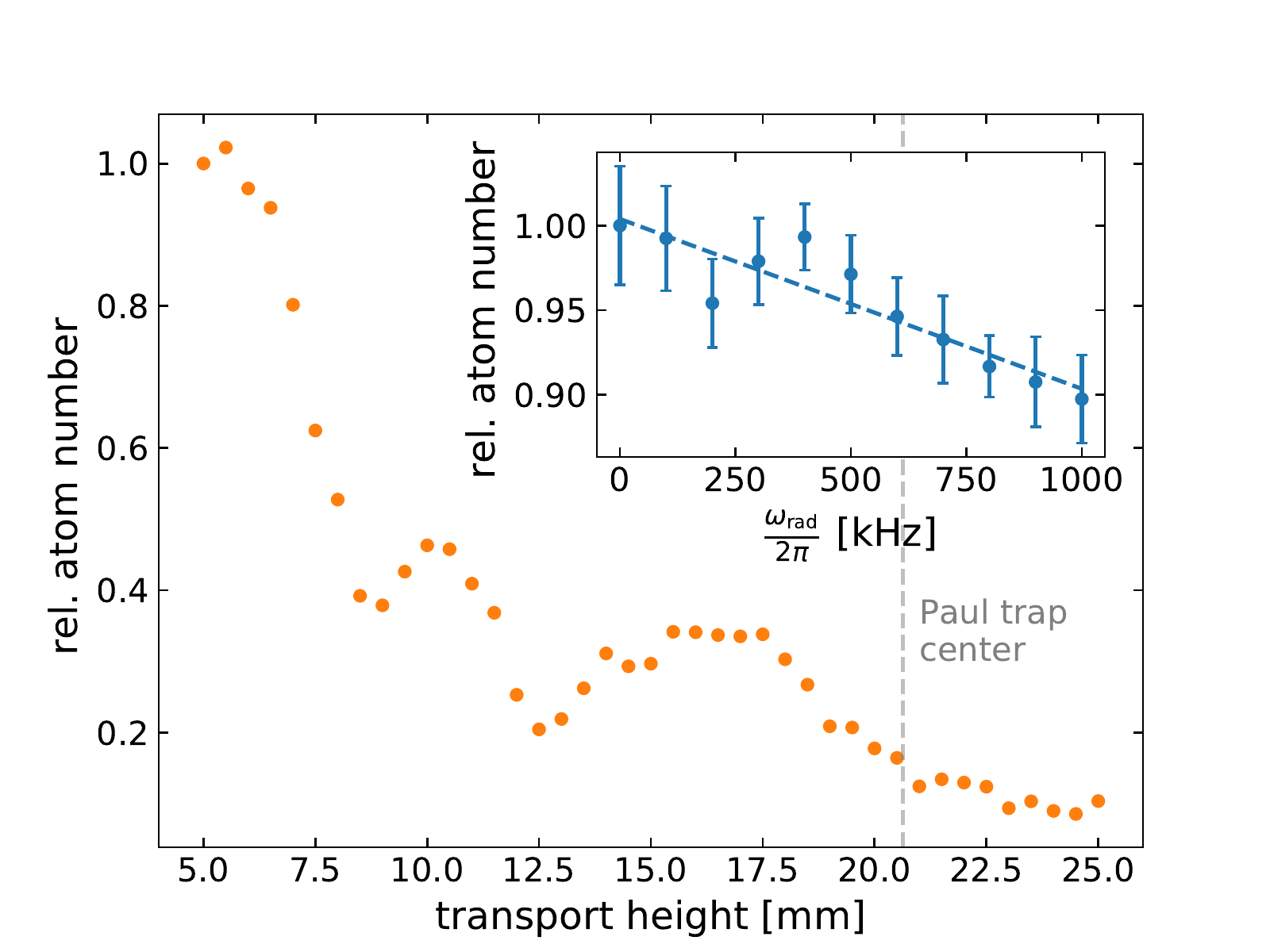}
	\caption{Atom loss during magnetic transport {from the mMOT, located at about $0\,$mm,} towards the Paul trap centered at $\approx21\,$mm in dependence of the transport height. The visible features can be related to the geometry of the MOT mirror and the blade electrodes.  Inset: Atom loss in dependence of the radial trapping frequency of the Paul trap for trapped Yb$^+$ ion. The trap frequency is directly proportional to the radio frequency voltage supplied to the Paul trap. We do not observe a significant increase in spin flip losses as we increase the radio frequency power.}
	\label{fig:RelAtomLoss}

\end{figure}

{In a next step the atoms are transported upwards within 120~ms into the ion trap by dynamic adjustment of the magnetic trapping field. During the transport we compress the atomic cloud by ramping the field gradient from $g_z=0.44$\,T/m to $g_z=2.8$\,T/m in order to prevent losses due to the geometric constrains of the ion trap. To investigate the effect of the geometry of our trap we measured the atom loss for different transport heights. This is depicted in Fig.~\ref{fig:RelAtomLoss} and shows that we have a notable atom loss of about 80\,\%. Furthermore, we see a temperature increase of roughly 30\% due to the magnetic transport. However, the temperature does not depend on the trap frequency of the Paul trap.

Another point of concern in our scheme is the occurrence of ion trap radio-frequency induced spin flips that would lead to atom loss since the atoms are pumped to high-field seeking states. In particular, our ion trap operates at a trap drive frequency of about $2\pi\times$2~MHz, such that a resonance condition occurs at a field of $\sim$~19~mT. For the maximum achieved magnetic field gradient, this corresponds to a distance of $\sim$~100~$\mu$m from the magnetic trap center and a potential energy of $\sim$~138~$\mu$K. The radio frequency knife caused by the ion trapping field thus cuts into the atomic cloud. We measure the remaining atom number after transporting the cloud into and back out of the ion trap for different radio frequency powers of the Paul trap, expressed in units of the corresponding radial trap frequency of the} ion {$\omega_\text{rad}$.} The results {are presented in Fig.~\ref{fig:RelAtomLoss} (inset). Atom loss due to possible spin flips inside the rf field of the Paul trap turns out to be rather small.}

%

The relevant laser orientations with respect to the Paul trap are shown in Fig.~\ref{fig:LaserBeams} b). Once transported to the center of the ion trap, {the phase space density of the atoms is about 10$^{-7}$}, which is too low to load enough atoms in the optical dipole trap (ODT) for efficient forced evaporative cooling. To increase phase space density, we apply another step of laser cooling in a six-beam MOT configuration, with laser light tuned close to the D2-line. We keep this secondary MOT on for 1~ms. Although the large atomic cloud is at this stage overlapped with the ions, we do not observe ion loss due to the laser excitation of the atoms. The reason is most likely that the atomic cloud is so dilute that collisions only occur very rarely within the 1~ms time window of the MOT. During this second MOT-stage we switch on the laser beams for the ODT at full power ($P_\text{dip} \approx 150\,W$). Thus the atoms are loaded into the optical dipole trap. At the end of the MOT stage we switch off the repumper in order to pump the population into the lower hyperfine manifold of ${F}=1/2$. With optimized mode-matching between the optical dipole trap and the MOT we load about 10$^6$ atoms per spin state {$F=1/2, m_F=\pm 1/2$} at a temperature of $T_\text{atom} = 350\,\mu$K.

\begin{figure}
	\includegraphics[width=1\columnwidth]{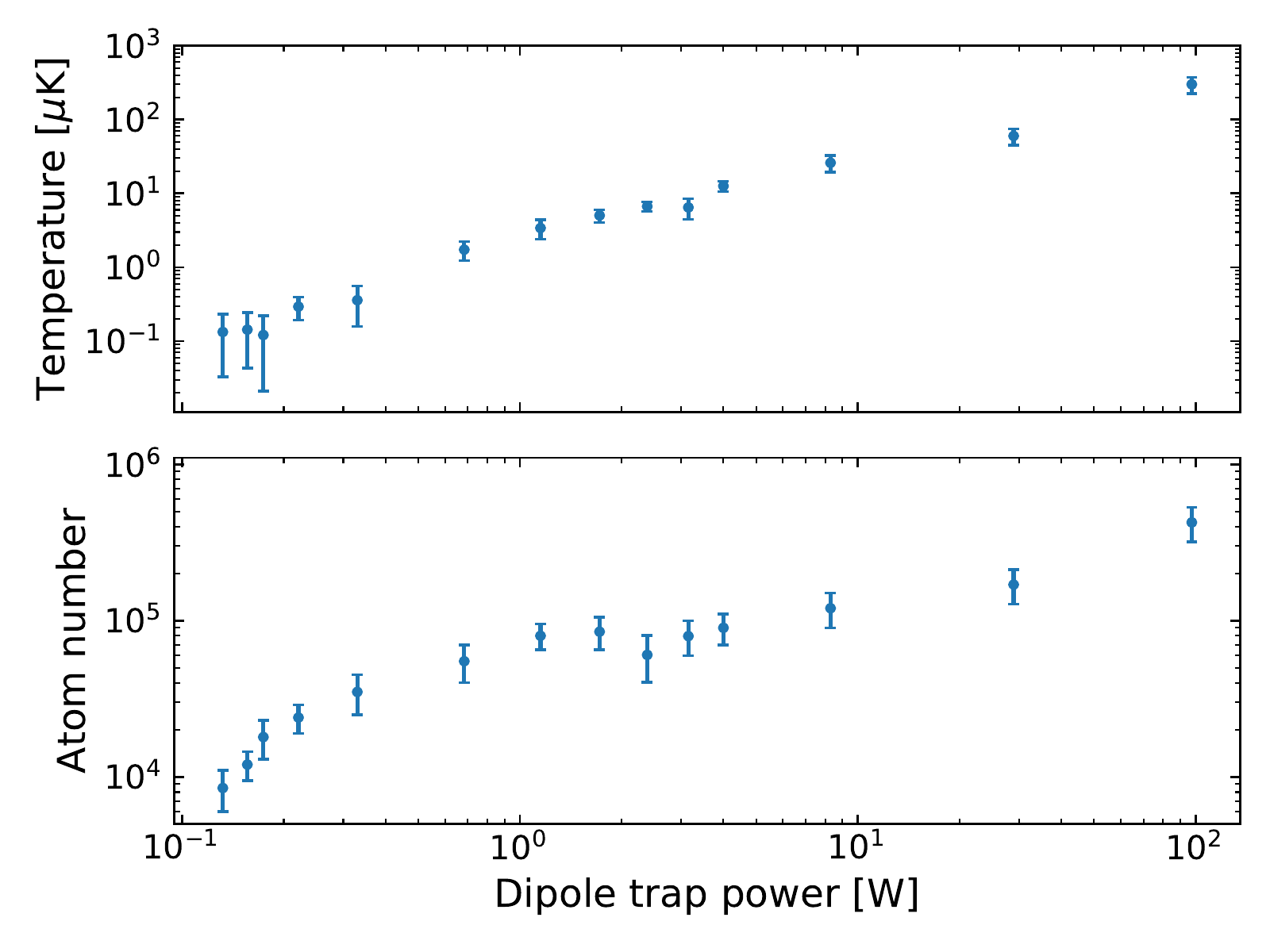}
	\caption{ Evaporative cooling of the spin mixture. Temperature (top) and atom number (bottom) in one of the spin states as a function of final power in the dipole trap. The data was obtained by absorption imaging and time-of-flight analysis. The full evaporation ramp takes 2~s. }
	\label{fig:EvapSummary}
\end{figure}

In order to reduce the temperature further, we employ evaporative cooling. The $F=1/2, m_F=\pm 1/2$ spin mixture in $^6$Li exhibits a broad Feshbach resonance around 83.2\,mT\,\cite{Zurn:2013}. By switching the polarity of the lower {Feshbach} coil, {we apply} a large homogeneous magnetic field of $B=78$\,mT corresponding to a scattering length of about $6000\,a_0$, with $a_0$ the Bohr radius. We ramp down the power of the dipole trap from an initial value of $P_{\text{dip}} = 150$\,W to a final value of $P_{\text{dip}} = 120$\,mW in 2\,s. The results of our evaporation ramp is shown in Fig.~\ref{fig:EvapSummary}. For the lowest temperatures, we get 10$^4$ atoms at $T_{\text{a}}=0.17(3)\,\mu$K. At low temperatures and in the vicinity of the Feshbach resonance, $^6$Li dimers are formed during evaporation. In order to prevent molecule formation, we can optionally switch the magnetic field to $B_0 = 30$\,mT before the last evaporation stage when the atoms reach a temperature of $T=15\,\mu$K. In this case the scattering length is reduced to $-300\, a_0$ at the expense of a slightly reduced atom number due to the less efficient final evaporation. 

The far-detuned light for the optical dipole trap (ODT) stems from a 200\,W fiber laser at $1070\,$nm. The output power of the laser can be controled in a range of 20\,W--200\,W. For further power reduction during evaporation and switching of the light we use an AOM in single pass configuration. Our ODT is set up in crossed beam configuration, with both beams propagating through holes in the endcaps of the Paul trap. The beams cross at an angle of 10$^\circ$ and have a minimal beam waist of $40\,\mu$m in the crossing point. A half-waveplate is used to provide lin-per{p}-lin polarization to avoid the creation of an optical lattice. To improve the mode-matching of the ODT with the upper MOT we employ a time-averaged potential. We modulate the AOM drive frequency using a triangle modulation signal at 4\,MHz and a modulation depth of 12~MHz from an arbitrary waveform generator. Due to the frequency dependence of the Bragg angle in the AOM this leads to a fast spatial modulation of the potential. On average this generates a shallower, broader potential, which results in better mode-matching with the upper MOT. During the first part of the evaporation we reduce the modulation to zero.

The atoms can either be imaged at the location of the initial magnetic trap, or once they have been transported into the Paul trap. In the latter case, we use absorption imaging along the axis of the Paul trap, which corresponds to the long axis of the optical dipole trap. In particular, the absorption beam is sent through holes in the endcap electrodes of the Paul trap as shown in Fig.~\ref{fig:LaserBeams}~b).
 Here, we have the option to image at a magnetic field of $\sim$~80~mT where the Paschen-Back effect allows us to detect each of the two spin states of the atoms independently. The imaging at the location of the Paul trap has a magnification of~2.2. 

\subsection{Overlapping atoms with ions}
\label{Subsec_overlap}
\begin{figure}
	\includegraphics[width=0.95\columnwidth]{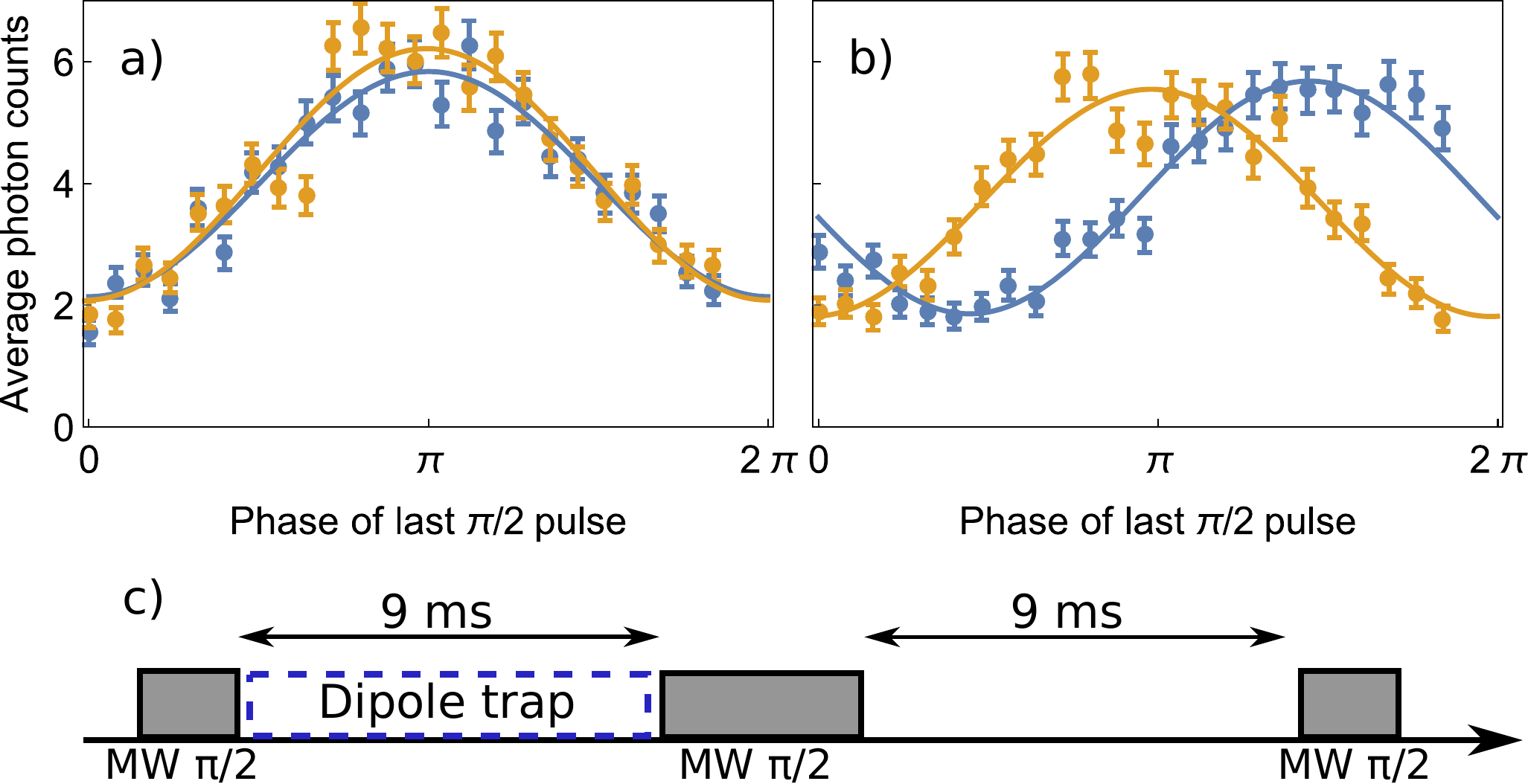}
	\caption{Dipole trap alignment to the trapped ion. Differential AC-Stark effect on the ion with (blue) dipole trap on and (orange) without dipole trap. a) For a misaligned laser the effect vanishes whereas for an overlapped beam b) an AC-Stark shift down to a few Hz is detectable. c) Microwave pulse sequence $\pi/2$ - $\pi$ - $\pi/2$ (spin-echo) with scanned phase on the second $\pi/2$ pulse, to measure the intensity from an incident beam between the first two pulses.
	}
	\label{fig:DipTrapAlignment}
\end{figure}

{Precise overlap of the dipole trap with the ion is essential for buffer gas cooling.}
\paragraph{Alignment of the dipole trap with the ion}
We maximize the overlap using the differential Stark shift of the 1070\,nm dipole trap beam on the $\ket{S_{1/2},F=0,m_F=0}\leftrightarrow\ket{S_{1/2},F=1,m_F=0}$ transition in $^{171}$Yb$^+$. We prepare the ion in the $\ket{S_{1/2},F=0,m_F=0}$ state before applying a $\pi/2$-pulse on the $\ket{S_{1/2},F=0,m_F=0}\leftrightarrow\ket{S_{1/2},F=1,m_F=0}$ transition. We switch on the dipole trap and wait for {9\,ms}. Subsequently we switch off the dipole trap, apply a $\pi$-pulse, wait another {9\,ms} and apply a final $\pi/2$-pulse. {To obtain Ramsey fringes, we scan the phase of the second $\pi/2$ pulse.} With this spin-echo sequence, see Fig.~\ref{fig:DipTrapAlignment}, we are able to measure differential Stark shifts introduced by the dipole trap with a precision of $\sim$~1 Hz. By scanning the control voltages of the piezo mirror mounts and repeating the measurement we maximize the Stark shift and thus the overlap of the dipole trap with the ion.

\paragraph{Alignment of the atomic cloud with the ion}

To fine-tune the overlap of the atom cloud with the ion, we adjust the piezo mirror mounts to optimize ion loss following collisions of Li and Yb$^+$ in the $P_{1/2}$ state. For this, we Doppler cool the ion during overlap with the atomic cloud after evaporation. Collisions result in charge transfer with high probability for populating the electronically excited states of Yb$^+$, which leads to ion loss~\cite{Joger:2017}.

\section{Ion spectroscopy}
\label{Sec_spectro}

\subsection{Thermometry}
\label{Subsec_thermo}

To study ultracold atom ion collisions it is important to accurately determine the average kinetic energy of the ion. We perform resolved sideband spectroscopy on the $S_{1/2} \leftrightarrow D_{5/2}$ quadrupole transition at 411\,nm wavelength.
With this, we can precisely determine the secular temperature of the ion and its micromotion. We prepare a $^{171}$Yb$^+$ ion in the $\ket{S_{1/2},F=0,m_F=0}$ ground state. We excite the first-order magnetic field-insensitive transition to the $\ket{D_{5/2},F=2,m_F=0}$ state. Choosing a magnetic field alignment at an angle of 45$^{\circ}$ with respect to the wavevector of the laser $\vec{k}_{411}$ and a polarization along the projection of the magnetic field in the plane of incidence, we maximize the transition strength on the $\Delta m_F = 0$ transition while minimizing the $\Delta m_F = \pm 1$ transition strength~\cite{Roos:2000}. 
The $D_{5/2}$ state with a lifetime of $\tau_D = 7.2\,$ms decays with 83\% probability to the long-lived $F_{7/2}$ state and with 17\%  probability back to the ground state~\cite{Taylor:1997}. We detect a successful excitation by illuminating the ion with Doppler cooling light at 369\,nm while also coupling the hyperfine levels of the ground state via microwave radiation at 12.64\,GHz and detecting the scattered light on a photomultiplier tube. An ion in the $F_{7/2}$ state scatters no photons while an ion in the $S_{1/2}$ state does. The long lifetime of the $F_{7/2}$ state ($\tau_F \approx 10\,$years) allows for basically arbitrarily long detection times. We choose a detection time of 100\,ms in order to achieve high-fidelity state detection, while not introducing too much delay in measurement time.
Note that due to the long detection time compared to the lifetime of the $D_{5/2}$ state and its branching fractions into the $F_{7/2}$ and $S_{1/2}$ states, we are limited to a maximal probability of 83\% to measure a dark ion.
	
From the excitation dynamics on the $S_{1/2} \leftrightarrow D_{5/2}$ quadrupole transition we can determine the	motional state of the ion. The Rabi frequency
$\Omega$ of oscillations on the spectroscopy transition depends on the population of
motional states with quanta $n_i$ in the secular motion of the ion~\cite{Leibfried:1996,Meir:2016},

\begin{equation}\label{Eq_Rabi}
	\Omega= \Omega_0\prod_{i=x,y,z} e^{-\eta_i/2}L_{n_i}(\eta_i^2).
\end{equation}

Here, $\Omega_0$ is the ground state Rabi frequency and $\eta_i=k_il^{\rm ho}$
the Lamb-Dicke parameter with $k_i$ the wavevector of the
411~nm light projected onto the direction of ion motion $i$ while $l^{\rm
ho}=\sqrt{\hbar/(2m_{\rm ion}\omega_i)}$ denotes the size of the motional ground state
wavepacket. The function $L_{n_i}(\eta_i^2)$ represents the Laguerre polynomial. We have $\eta_x=\eta_y=\eta/\sqrt{2}$ and $\eta_z=0$ for the measurements on the radial motion, since the laser has a 45$^{\circ}$ angle with respect to the $x$- and $y$-direction for these measurements.

The dependence on the motional state in Eq.~\ref{Eq_Rabi} results in mixing of Rabi frequency
components when the ion is not in the ground state of motion. From the damping rate of the Rabi flops we can infer the occupation of harmonic oscillator states, from which we can determine the average ion energy. We fit the measured probability  to be in the $S_{1/2}$ state $p_{S}$ as a function of the pulse duration $\tau_{411}$ to a model that assumes a thermal distribution with $P_{\bar{n}_{x,y}}(n)=\bar{n}_{x,y}^n/(1+\bar{n}_{x,y})^{n+1}$ for each direction of motion $x$ and $y$, assuming $\bar{n}_x$=$\bar{n}_y$

\begin{equation}\label{Eq_Fit_nbar}
p_{S}= 0.585+ \frac{0.83}{2}  \sum_{n_x,n_y} P_{\bar{n}_{x}}(n_x) P_{\bar{n}_{x}}(n_y)\cos\left(2\Omega \tau_{411} \right).
\end{equation} 

The ion temperature (in its secular motion) is given by $T_{\text{sec}}^{\perp}=\hbar\omega(\bar{n}+1/2)/k_\text{B}$ with $\bar{n}=(\bar{n}_x+\bar{n}_y)/2$ the average number of motional quanta. The factors 0.83 and $0.585=(1-0.83/2)$ arise due to the branching ratio of 0.83 from the $D_{5/2}$ state to the $F_{7/2}$ state~\cite{Feldker:2020}.

 In our system we did not find any deviations from a thermal energy distributions for the ion~\cite{Feldker:2020}. We attribute this to the large mass ratio where the modifications to the distribution do not seem to play a role~\cite{Rouse:2017,Meir:2016}. Note that for smaller mass ratios, non-thermal energy distributions have been studied extensively for buffer gas-cooled ions in a Paul trap~\cite{DeVoe:2009,Zipkes:2011,Chen:2014,Rouse:2017,Weckesser:2015,Meir:2016}.



{The described method gives the most reliable results in the Lamb-Dicke regime where $\eta \sqrt{2\bar{n}+1} \ll 1$.  Particularly, in the axial direction of the trap we have low trapping frequencies which makes it challenging to enter the Lamb-Dicke regime. Thus we use an alternative thermometry method for which we measure the envelope of the sideband spectrum. From the broadening of the transition we determine the average speed of the ion and thus the temperature. While this method does not rely on the Lamb-Dicke regime and thus works for shallow traps or higher temperatures, it is less exact due to a variety of other sources of line broadening. The temperature of the ion is related to the standard deviation of the Gaussian spectral distribution in Hz, $\sigma_{\text{spec}}$, by $T_{\text{sec}}^{\text{ax}}=m_\text{ion}\lambda_{411}^2\sigma_{\text{spec}}^2/k_\text{B}$. }

\subsection{Micromotion compensation}
\label{Subsec_MMcomp}

Precise micromotion determination and compensation is crucial for buffer gas cooling to ultracold temperatures. We use a set of complimentary methods that is also partially described in Ref. \cite{Joger:2017,Fuerst:2018,Feldker:2020} to accurately measure the micromotion of the system. 

Three types of excess micromotion are generally distinguished~\cite{Berkeland:1998}. First, unwanted static electric fields push the ion out of the center of its trap such that it experiences a non-zero radio frequency field. We refer to this type of micromotion as radial micromotion. Secondly, excess micromotion may occur due to a phase difference between the supplied rf voltages on the blades. This type of micromotion is known as quadrature micromotion. This effect can occur e.g., because of length differences in the high-voltage cables and connectors. Note that this is more likely to cause problems in Paul traps that are driven at high frequencies. Finally, imperfections in the Paul trap may cause oscillating fields along its trap axis. This we refer to as axial micromotion. In our trap, the oscillating fields along the trap axis are inhomogeneous, such that there is a point in space where the axial micromotion is minimal. Note however, that even if the axial oscillating fields would exactly vanish in this point, they still lead to excess micromotion for ion crystals. This configuration can be quantified by introducing an axial stability parameter $q_\text{ax}$. In our setup, {we estimate that $q_\text{ax}/q_\text{rad}\approx$~10$^{-2}$~\cite{Joger:2017},  {where $q_\mathrm{rad}$ is the radial stability parameter}}

\paragraph{Radial micromotion compensation}

As shown in Fig.~\ref{fig:SetupVert}, our trap features two pairs of dedicated electrodes for the compensation of stray electric fields. It turned out that the trap effectively shields any field we apply in the horizontal direction, so that we cannot use these electrodes to compensate for stray electric fields in this direction. However, due to small imperfections in the trap manufacturing and charges accumulating on it over time, there is a small dependence of the electric field in the radial directions on the endcap voltage. For a full compensation of stray electric fields we first compensate the field in the horizontal direction by supplying appropriate voltages to the endcap electrodes. This introduces an undesired stray field in the vertical direction which we subsequently compensate by applying the appropriate voltages to the compensation electrodes. 

While this scheme allows for a complete compensation of stray electric fields it has the disadvantage that the axial trap frequency cannot be chosen freely. What is more, we observe that the ideal voltage setting on the endcap electrodes changes on the timescale of weeks, such that we require increasingly larger voltages. We remedy this by regularly applying intense heating pulses of a few seconds to one side of the endcap with a high-power (30~W) infrared laser. In this way we can modify the charge distribution and thus shift the endcap voltages required for compensation to the desired axial trap frequencies. After this treatment the electric fields drift for a few hours, but remain stable on the timescale of weeks afterwards. Thus we have to apply the heat treatment of the endcaps {only if we want to significantly} change the trap settings.

\begin{figure}
	\includegraphics[width=\columnwidth]{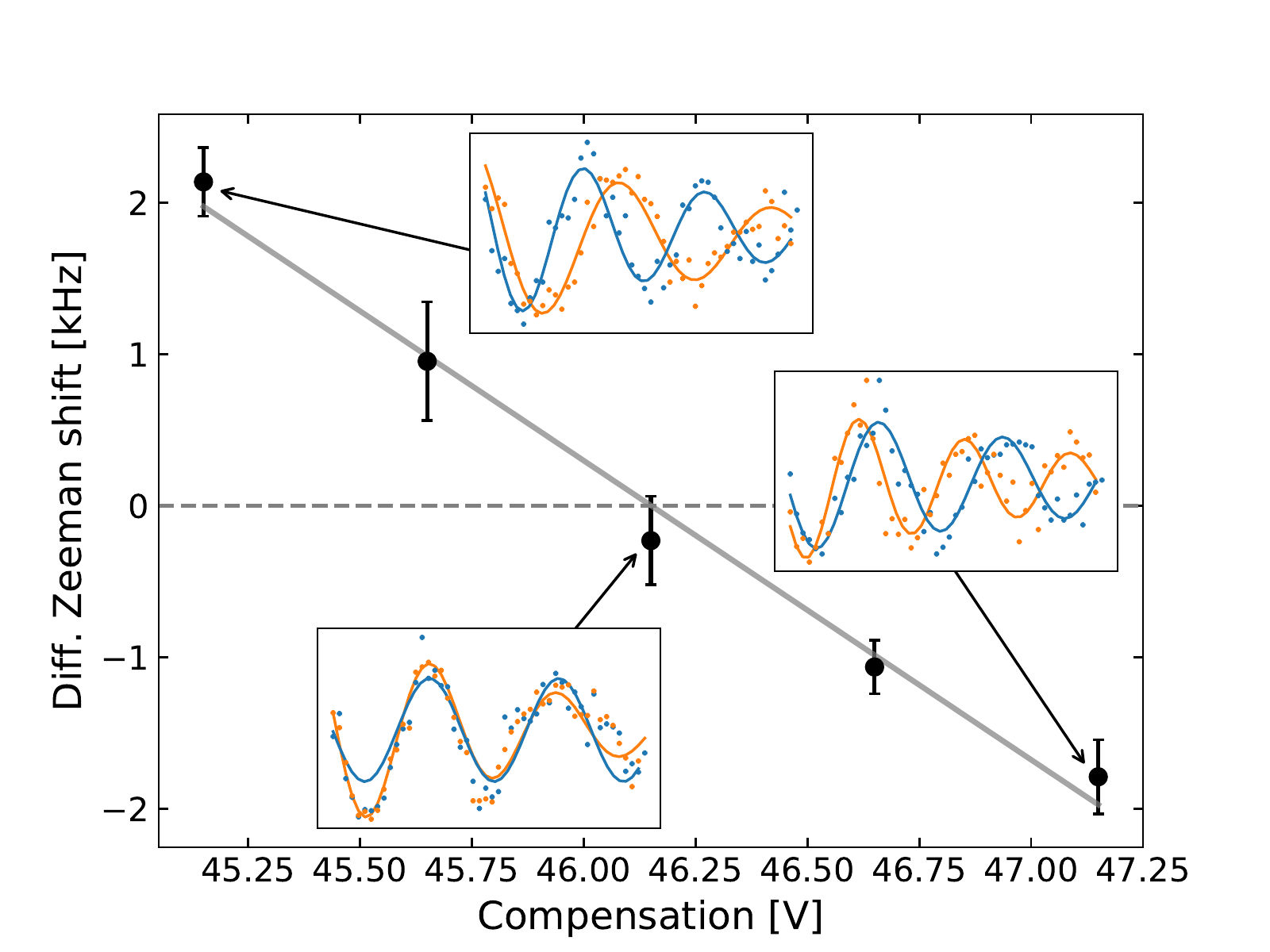}
	
	\caption{Differential Zeeman shift versus compensation electrode voltage with linear fit (gray line). Insets show Ramsey fringes for a low (blue) and high (orange) trap frequency for two uncompensated cases (right \& top) and one compensated case with vanishing shift (bottom).} 
	\label{fig:RamseyMicromotion}
\end{figure}

In the horizontal direction, we determine the stray electric fields $E_{\text{dc}}$ by measuring the position shift of the ion with the camera. The ion position is given by:

\begin{equation}\label{Eq_xvsEdc}
x(\omega_{\text{rad}}) = E_{\text{dc}} \frac{e}{\omega_{\text{rad}}^2 m_{\text{ion}}},
\end{equation}

\noindent where $e$ is the elementary charge. 
The resolution of our imaging system allows us to determine the average position of the ion with a precision of about 200\,nm. We can lower the trap frequency to about $\omega_\text{rad} = 2\pi \times 20\,$kHz without losing the ion. With these settings, we can compensate fields to $E_{\text{dc}} \sim 10$\,mV/m. The drift during a full day of measurements is typically $\Delta E_{\text{dc}} < 50\,$mV/m.

In the vertical direction we use the frequency shift of the $\ket{S_{1/2}, F=0} \leftrightarrow \ket{S_{1/2}, F=1,m_F=1}$ microwave transition in a magnetic field gradient to determine the position shift. {The results are shown in Fig.~\ref{fig:RamseyMicromotion}.} We use a Ramsey type experiment at trap frequencies of $\omega_{\text{rad,low}}= 2\pi \times 99\,$kHz and $\omega_{\text{rad,hi}}= 2\pi \times 205\,$kHz with variable wait time during the $\pi/2$-pulses (insets of Fig.~\ref{fig:RamseyMicromotion}). In particular, the relative position shift of the ion due to a static offset field is given by

\begin{equation}\label{Eq_RamseyMM_det}
\Delta r_\text{v}\approx \frac{e E_\text{rad,v}}{m_\text{ion}}\left(\frac{1}{\omega^2_{\text{rad,hi}}} - \frac{1}{\omega^2_{\text{rad,low}}} \right),
\end{equation}  

\noindent which can be related to the frequency shift between the two measurements:

\begin{equation}
\Delta f_\text{mw} = g_\text{v} \frac{\mu_\text{B}}{2\pi\hbar}\Delta r_\text{v}.
\end{equation}

\noindent Here, $\mu_\text{B}$ denotes the Bohr magneton and the magnetic field gradient is set to $g_\text{v}=134\,$mT/m. We measure the dc electric field with a precision of $E_\text{dc} \sim 20\,$mV/m, while day to day variations are $E_\text{dc} < 60\,$mV/m. However, the two methods described above provide an indirect measurement of the micromotion induced by stray electric fields only. Other types of micromotion such as quadrature micromotion due to a phase shift of the rf-signal on the two rf-electrodes are not detected. 

A direct micromotion measurement relies on sideband spectroscopy of the $S_{1/2} \leftrightarrow {D}_{5/2}$ transition at 411\,nm wavelength. The transition strength on the micromotion sideband is directly related to the micromotion amplitude~\cite{Berkeland:1998}:

{
\begin{equation}\label{Eq_MMJ}
\frac{\Omega_\text{sb}}{\Omega_\text{{car}}}=\frac{J_\text{1}(\beta_\text{mm})}{J_\text{0}(\beta_\text{mm})},
\end{equation}
}

\noindent
{with $\Omega_\mathrm{car}$ and $\Omega_\mathrm{sb}$ denoting the Rabi frequency of the carrier and sideband respectively, {and $\beta_\mathrm{mm}$ is the modulation index which can be equated to the wavevector $\vec{k}_{411}$ in direction of ion motion~$\beta_\text{mm} = \vec{k}_{411}\cdot\vec{r}_\text{mm}$ with $|\vec{r}_\text{mm}|$ the micromotion amplitude and $J_\text{1}(\beta_\text{mm})$ and $J_\text{0}(\beta_\text{mm})$ Bessel functions of the first kind.}
We deduce the micromotion amplitude from the measured Rabi frequencies $\Omega_{\text{car}}$ and $\Omega_{\text{sb}}$ at laser powers $P_{\text{car}}$ and $P_{\text{sb}}$. The measured value indicates the projection of the total micromotion on the $k$-vector of the interrogation beam. 

We use this method to calibrate the field $E_\mathrm{dc}$ versus voltage on the compensation
electrodes by comparing
the Rabi frequency on the micromotion sideband and the carrier at
V$_{\text{comp}} = $7\,V, and setting $\omega_{\text{rad}}=2\pi$~330~kHz. The results are shown in Fig.~\ref{fig:MMCalibration} and we find $\Omega_{\text{sb}}=2\pi \times
28.3(0.9)\,$kHz and $\Omega_{\text{car}}=2\pi \times 39.0(1.2)\,$kHz. 
This yields a modulation index
$\beta_{\text{mm}} = e E_\text{dc} k_{411} q/(2m_\text{ion}\omega_\text{rad}^2)= 1.18(5)$, from which we obtain a scaling of $E_\text{dc}=0.34(2)\times V_\text{comp}/\text{V}$~V/m.

After carefully compensating all stray electric fields, we compare the Rabi frequency on
the carrier, $\Omega_{\text{car}}=2\pi \times 32.0(0.8)\,$kHz at a laser power
of $P_{\mathrm{car}} = 32\,\mu$W with the Rabi frequency on the micromotion sideband
$\Omega_{\text{sb}}=2\pi \times 7.0(0.5)\,$kHz at $P_{\mathrm{sb}} = 840\,\mu$W. This corresponds to a modulation index of $\beta_{\text{mm}}=0.085(10)$, which we attribute to the remaining radial micromotion and quadrature micromotion.

\begin{figure}
	\includegraphics[width=0.95\columnwidth]{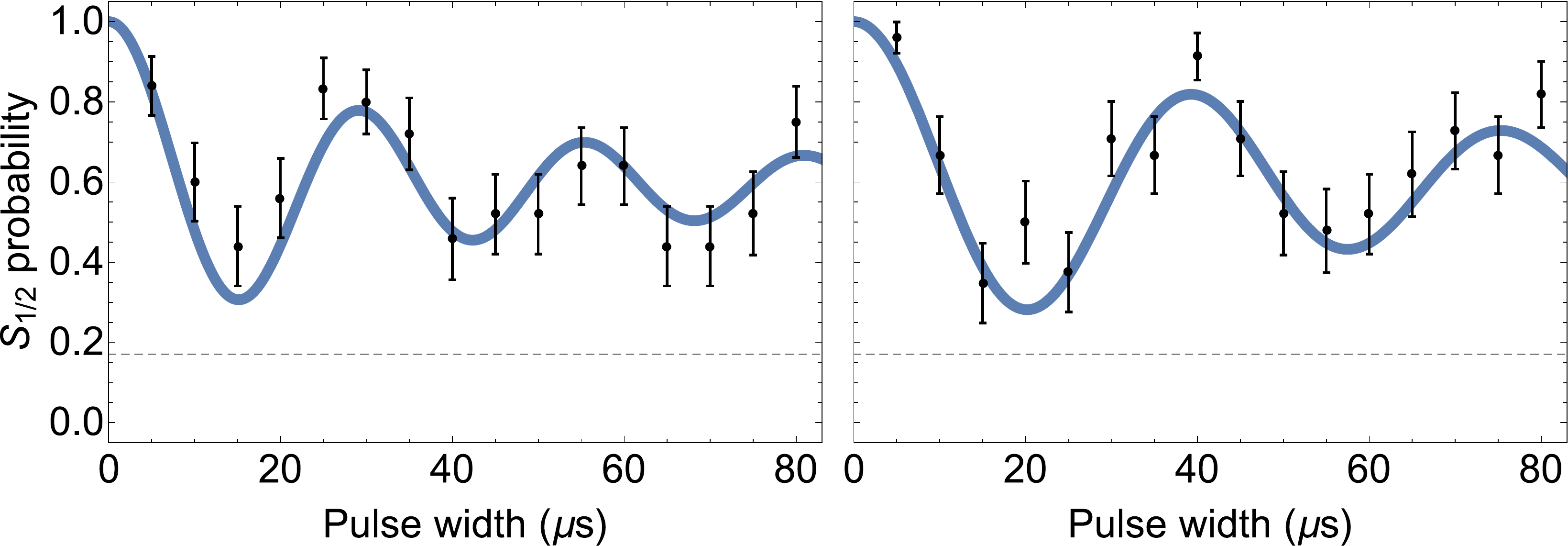}
	\caption{ Calibration of the offset field $E_\text{dc}$ for radial micromotion compensation.  Rabi flops on the 411~nm transition after buffer gas cooling, on a) the carrier and b) the sideband for a compensation voltage of $V_\text{comp}=$~7~V as compared to the optimal micromotion compensation. From the ratio, we can obtain the scaling of the supplied electric field $E_\text{dc}$ and compensation voltage  $V_\text{comp}$ as explained in the text. {Due to the branching ratio of the $D_{5/2}$ state, a maximal contrast of $83\%$ can be measured.}}
	\label{fig:MMCalibration}
\end{figure}

\paragraph{Axial micromotion.}

Finite size effects of the linear Paul trap lead to rf-electric fields in the
direction of the trap axis that only disappear in one point along the trap axis. We position the single ion in our trap to this point and
measure an upper limit to the remaining axial micromotion by comparing a frequency scan over the carrier
at very low power $P_{\mathrm{car}} = 61\,\mu$W with a scan over the micromotion sideband
at full power $P_{\mathrm{sb}} = 21.7\,$mW~\cite{Feldker:2020}. We calculate an upper bound
to the axial oscillating field of 
$<1.6$~V/m~\cite{Feldker:2020}. 

\begin{figure}
	\includegraphics[width=0.95\columnwidth]{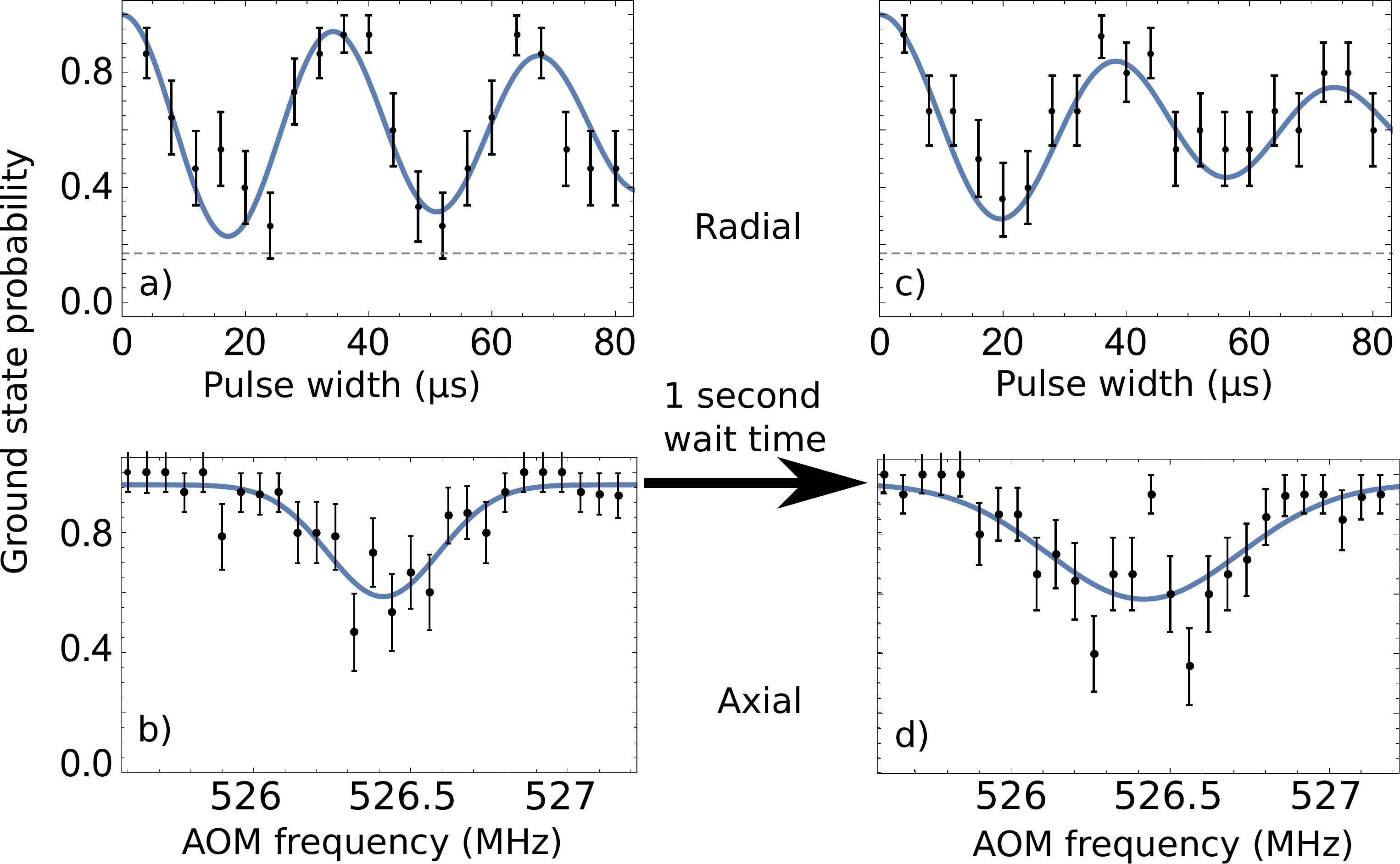}
	\caption{Kinetic energy measurements of the ion after buffer gas cooling [a) and b)] and after an additional $1\,$s waiting time [c) and d)]. a) After buffer gas cooling, we measure Rabi flops on the $S_{1/2} \leftrightarrow D_{5/2}$ quadrupole transition. We fit the flops to our model Eq.~(\ref{Eq_Fit_nbar}) to obtain the average occupation number of motional quanta $\bar{n}$ and thereby the secular temperature of the ion $T_{\text{sec}}^{\perp}$, as explained in the text. b) In the axial direction, we instead study the spectral width of the laser excitation to obtain the ion temperature $T_{\text{ax}}^{\perp}$. On the $x$-axis, is given the frequency supplied to an AOM to control the frequency of the laser with precision at the kHz level. c) and d) show the same type of measurement as a) and b) but after waiting another 1~s without any atoms present. In this way, we can obtain the background heating rate of the ion as explained in the text. }
	\label{fig:HeatingRates}
\end{figure}

\section{Buffer gas cooling}
\label{Sec_atomsandions}

In this section we, describe experiments on interacting atoms and ions. For these, we prepare an atomic cloud of about 2~$\times$~10$^4$ atoms at a temperature of $T_{\text{atom}}$=10~$\mu$K in a balanced spin mixture of the lowest energetic states, $|F=1/2,m_F=\pm1/2\rangle$ at a density of $\rho_{\text{atom}}\sim$~31(15)~$\times$~10$^{15}$/m$^3$. The cloud is held 200~$\mu$m below a single trapped and laser-cooled $^{171}$Yb$^+$ ion, prepared in its lowest energy state, $|F=0,m_F=0\rangle$. As a next step, we transport the cloud up to the ion and let the system interact for 1$\,$s, after which we release the atoms by switching off the dipole trap and interrogate the ion with our 411~nm spectroscopy laser as described above. The magnetic field is set to 0.4~mT during the overlap of the ion and atoms.  

The measurements of the ion's kinetic energy in its secular motion is shown in Fig.~\ref{fig:HeatingRates} a) and b) for axial and radial direction, respectively. From fitting the data after buffer gas cooling by the ultracold Li cloud, we find an average number of motional quanta $\bar{n}=5.3(1.8)$ in the radial directions of motion, corresponding to $T_{\text{sec}}^{\perp}$=90(35)~$\mu$K, or about a factor 5 below the Doppler cooling limit of Yb$^+$. For the axial direction, we obtain the ion's temperature from the spectral width of the excitation as described above. Here, we find $T_{\text{sec}}^{\text{ax}}$=108(25)~$\mu$K. Immersing the ion into the cloud for 1~s and repeating for 5000 times, we infer an ion loss rate of $\lesssim$10$^{-4}\times\gamma_\text{L}$, in agreement with calculations~\cite{Tomza:2015}.

\subsection{Buffer gas cooling and heating rates}
\label{Subsec_heatingrates}

To investigate whether background heating due to electric field noise may pose limitations to our buffer gas cooling, we release the atoms and wait for 1~s to see whether the buffer gas cooled ion heats up. The results are shown in Fig.~\ref{fig:HeatingRates} c) and d) for axial and radial direction, respectively. After the waiting time, we obtain $\bar{n}=10.6(2.1)$ and a wider spectrum for the axial motion. {From these measurements, we extract a heating rate  of 85(50)~$\mu$K/s in the radial direction and 197(48)~$\mu$K/s in the axial direction in the absence of the buffer gas. We attribute the higher heating rate in the axial direction to the lower confinement in that direction~\cite{Brownnutt:2015}.}

The background heating competes with the buffer gas cooling and leads to a larger final temperature of the ion. The effect can be considered via a rate equation:

\begin{equation}\label{Eq_rate}
\frac{dT_\text{sec}(t)}{dt} = -\gamma_\text{cool}
(T_\text{sec}(t)-T_{\infty})+\Gamma_\text{heat},
\end{equation}
\noindent
where $T_\infty$ is the equilibrium temperature.  Here, $\Gamma_\text{heat}$ is the background heating rate, which is independent of its motional state in the ultracold regime considered~\cite{Brownnutt:2015}. 
At equilibrium, we see that the background heating increases the final temperature of the buffer gas cooled ion by $\Delta T=\Gamma_\text{heat}/\gamma_\text{cool}$. With $\Gamma_{\text{heat}}=85(50)$~$\mu$K/s and $\gamma_{\text{cool}}$=1/244(24)~ms$^{-1}$ as measured in our experiment~\cite{Feldker:2020}, we get $\Delta T_{\text{sec}}^{\perp}=$~21(12)~$\mu$K for the radial direction. For the axial direction, we find $\Delta T_{\text{sec}}^{\text{ax}}= 48(13)$~$\mu$K, assuming the same $\gamma_{\text{cool}}$ as for the radial direction. Combining these results and comparing them to the data suggests $T_{\infty}^{\perp}=$~69(37)~$\mu$K while $T_{\infty}^{\text{ax}}=$~60(28)~$\mu$K in agreement with thermalization between the directions of motion in the absence of background heating.

Classical molecular dynamics simulations that neglect the background heating give as a result $T_{\infty}^{\text{ax}}=T_{\infty}^{\perp}=$~38.2~$\mu$K~\cite{Fuerst:2018,Feldker:2020} and  strictly exponential thermalization dynamics, in agreement with~Eq.~(\ref{Eq_rate}). Note that these simulations use as input parameters the experimental settings as measured in the laboratory, including the measured levels of excess micromotion. One explanation for the slight discrepancy between the measured and simulated temperatures could be an overestimation of the ion energy in the experiment. This can happen as sources of loss of contrast in the Rabi flops, such as laser linewidth and power fluctuations are disregarded in the analysis to get the most conservative measurement of the ion temperature. 

It should be relatively straightforward to (locally) increase the density of the buffer gas, and thereby $\gamma_{\text{cool}}$, by an order of magnitude or more, by e.g. stronger optical confinement. This could eliminate the background heating as a limitation in buffer gas cooling altogether. Another option is to reduce the background heating rate. This may be feasible by noise reduction, however there may not be a lot to gain as the measured value of a few motional quanta per second is quite typical for the type of Paul trap employed~\cite{Brownnutt:2015}.  


\subsection{Excess micromotion and buffer gas cooling}
\label{Subsec_EMMbuffergascool}

\begin{figure}
	\includegraphics[width=0.95\columnwidth]{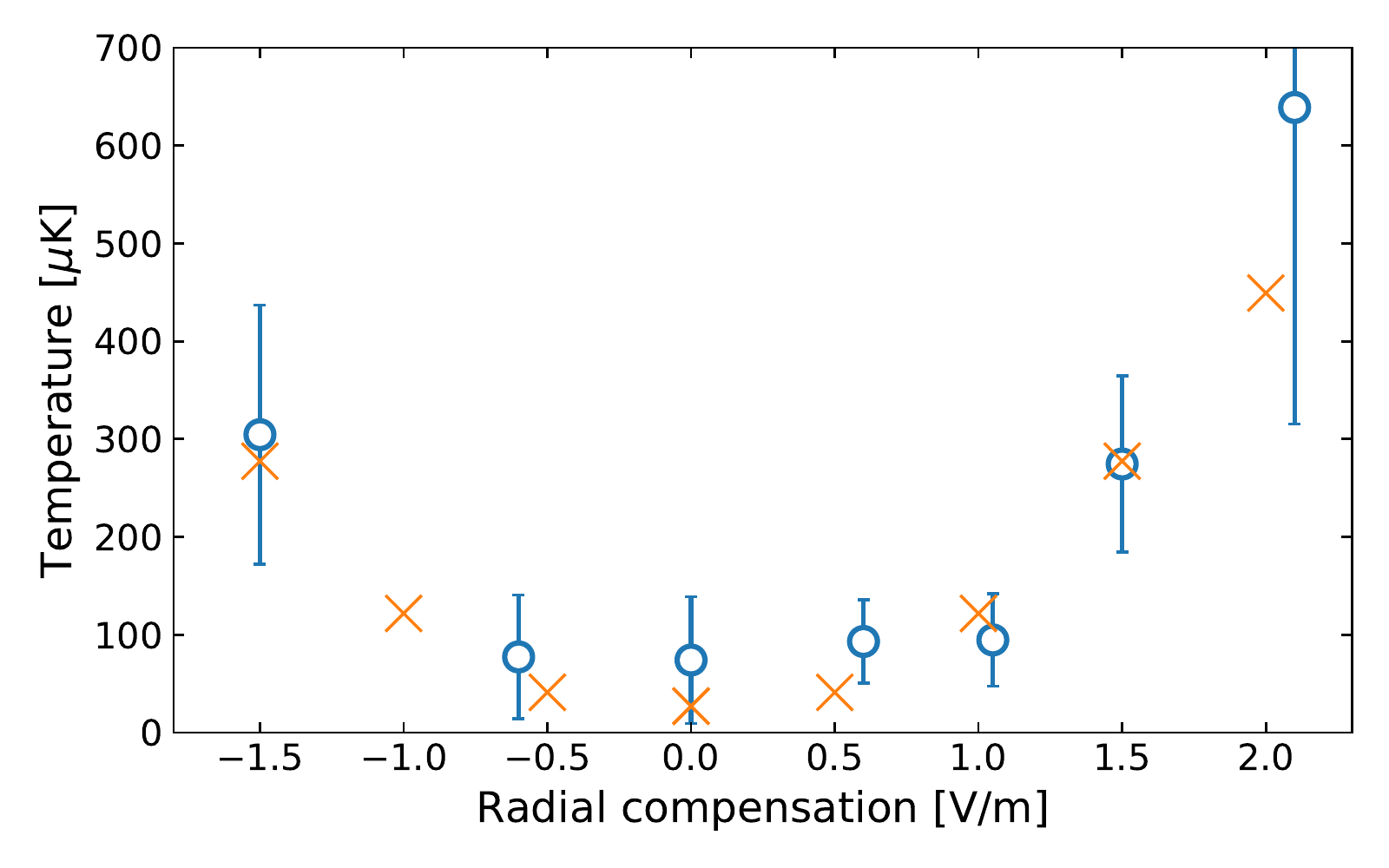}
	
	\caption{Ion temperature versus offset electric field causing excess micromotion obtained from measurement (circles) and from molecular dynamics simulation (crosses) with parameters as used in the experiment. }
	\label{fig:TvsComp}
\end{figure}

The effect of excess micromotion on the buffer gas cooling of a trapped ion has been studied in a number of experimental and theoretical works~\cite{Zipkes:2011,Harter:2013,Rouse:2018,Meir:2016}. It has been found that excess micromotion leads to much higher final ion temperatures than when considering an ideal Paul trap.  Numerical simulations reveal that $\gamma_{\text{cool}}$ is also weakly affected by excess micromotion, leading to a slight increase~\cite{Fuerst:2018}. The effect of excess micromotion in an ion-atom combination with a large mass ratio is of particular importance as it could have an impact on the prospects for reaching deep into the quantum regime of atom-ion interactions~\cite{Fuerst:2018}. 

In this section, we study the achievable ion temperatures when we buffer gas cool the ion while giving it a controlled amount of excess micromotion. To this end, we apply a dc offset voltage to a pair of compensation electrodes generating a field $E_{\text{dc}}$ that pushes the ion out of the center of the trap by a distance $x_{\text{EMM}}=eE_{\text{dc}}/(m_\text{ion}\omega_x^2)$. This gives the ion an excess micromotion amplitude of  $x_{\text{EMM}}q_\text{rad}/2$. In these experiments, we buffer gas cool the ion for 1~s, after which it is interrogated by the 411~nm laser to obtain its secular temperature  for each voltage settings, as described in Sec. \ref{Subsec_MMcomp}. 

{Our result shows that buffer gas cooling remains effective as compared to Doppler cooling up to an offset field of about 2~V/m. This is shown in Fig.~\ref{fig:TvsComp}, which depicts the effect of excess micromotion on the  ion temperature after buffer gas cooling for experimental measurements (circles).}
We compare the data to molecular dynamics simulations (crosses), which explain the data very well. In the simulation, we extract the ion's secular temperature by subtracting the theoretical energy for the used excess micromotion parameters \cite{Berkeland:1998} from the total average kinetic energy. More details on the molecular dynamics simulation can be found in Ref.~\cite{Fuerst:2018}.


 We conclude that we can control the excess micromotion of the trapped ion  and can predict the final temperature of a buffer gas-cooled ion under the influence of excess micromotion.

\section{Conclusion}
\label{Sec_conclusions}
In this paper, we have described an apparatus for studying interactions between ultracold atoms and laser-cooled, trapped ions. We have described the preparation of ultracold atomic clouds and their overlap with a single ion. We have shown how laser spectroscopy on the ion can be used to determine its kinetic energy after interacting with the atoms. We have presented data showing that the temperature of the secular motion of the ion reaches 90(35)~$\mu$K in the radial and 108(25)~$\mu$K in the axial direction, respectively, after buffer gas cooling with the ultracold atoms. Measurements without the atom cloud show a background heating rate of 85(50)~$\mu$K/s in the radial and 197(48)~$\mu$K/{s} in the axial direction due to electric field noise. The competition between this heating rate and the buffer gas cooling limits the attainable ion temperatures but significant improvements should be possible by increasing the density of the gas. 

We have presented our methods for detecting and compensating excess micromotion in the ion. We have measured the attainable temperatures in the secular motion of the ion under the influence of excess micromotion. The measured temperatures can be accurately reproduced using classical molecular dynamics simulations. 

We identify a number of future research directions in our system. In particular, the collision energy between the atom and ion reaches a regime where quantum effects are to be expected~\cite{Feldker:2020}. This opens up the possibility to find Feshbach resonances between the atoms and ions~\cite{Tomza:2015}. The system may also be viewed as a single charged impurity that is interacting with a fermionic bath. It will be particularly interesting to study this system in the quantum degenerate regime, where we can tune the bath from a Bose- Einstein condensate of weakly bound Li$_2$ dimers, to a degenerate Fermi gas using the broad Feshbach resonance at 83.2~mT~\cite{Julienne:2010,Schirotzek:2009,Cetina:2016}. For this we would have to increase the density of the gas which should be feasible by adding a dimple potential. We will investigate whether it is possible to  buffer gas cool a trapped ion (close) to its ground state of motion. In this regime, it would be possible to study the dynamics of non-classical states of ion motion and decoherence in a quantum bath~\cite{Leibfried:1996,Daley:2004,Krych:2013}. Finally, the buffer gas cooling and interactions in ion crystals~\cite{Bissbort:2013,Fuerst:2018} may be investigated. 



\section*{Acknowledgements}
We gratefully acknowledge support from the group of Selim Jochim in Heidelberg, Germany, during the construction phase of the experimental setup. 
This work was supported by the Netherlands Organization for Scientific Research
(Vidi Grant 680-47-538, Start-up grant 740.018.008 and Vrije  Programma 680.92.18.05)  and the European Research Council (Starting Grant No. 337638) (R.G.). R.S.L. acknowledges funding from the European Union’s Horizon 2020 research and innovation programme under the Marie Sklodowska-Curie grant agreement No 895473.

\section*{References}


%

\end{document}